# Distance Priority Based Multicast Routing in WDM Networks Considering Sparse Light Splitting

Fen ZHOU, Miklós MOLNÁR
Department of Computer Science
INSA de Rennes / IRISA
Rennes, France
{fen.zhou, molnar}@irisa.fr

Bernard COUSIN
University of Rennes I / IRISA
Rennes, France
bernard.cousin@irisa.fr

*Abstract*—As we know, the Member-Only algorithm in [1] provides the best links stress and wavelength usage for the construction of multicast *light-trees* in WDM networks with sparse splitting. However, the diameter of tree is too big and the average delay is also too large, which are intolerant for QoS required multimedia applications. In this paper, a distance priority based algorithm is proposed to build light-trees for multicast routing, where the *Candidate Destinations* and the *Candidate Connectors* are introduced. Simulations show the proposed algorithm is able to greatly reduce the diameter and average delay of the multicast tree (up to 51% and 50% respectively), while keep the same or get a slightly better link stress as well as the wavelength usage than the famous Member-Only algorithm.

## I. INTRODUCTION

Multicast communication is becoming increasingly important in the recent years because of its efficient resources usage and the increasing popularity of the point-to-multipoint multimedia applications. A multicast session typically involves a source and a set of destinations. In traditional data networks, usually, a multicast tree rooted at the source is constructed with branches spanning all the destinations to accommodate a multicast session. In order to support multicast data in WDM optical networks, a *light-tree* concept was proposed in [2], which is a tree in the physical topology and occupies the same wavelength in all the fiber links in the tree. The optical switch nodes in the tree should be able to transmit one incoming optical signal to several outgoing ports, which is called splitting capability. Note that optical switches with splitting capability are always much more expensive to build than those without due to their required components and complicated structure. Consequently, only a few nodes in the all optical network are splitters. This is characterized as *sparse light splitting*. Moreover, the wavelength continuity constraint should be also respected in lack of costly wavelength converters. Hence, multicast routing in WDM optical networks is greatly different from that in the electrical domain and one must consider the constraint on the capabilities of nodes in practical optical networks. Due to these physical constraints, supporting multicast routing in all optical network is both an important and a challenging work.

The hardness of the multicast routing in WDM networks with sparse light splitting capability has been discussed in many papers [1, 4~10] and different multicast tree or forest formation algorithms have been proposed. The main objective of this problem is to optimize the optical network resources in face of constraints. To solve this problem, the famous Steiner tree is always used, which is known to be NP-complete [3], when the multicast group has more than two members [4]. However, to build a Steiner *light-tree* with the minimum costs for a multicast session in all optical networks with optical constraints is even harder. So, heuristics should be used. In [1], four heuristics: Reroute-to-Source, Reroute-to-Any, Member-First and Member-Only are proposed. Among them, the well-known Member-Only algorithm is currently thought to have the best link stress and usage of wavelength channels in all optical networks with sparse light splitting and without wavelength conversion. It is a modified version of Takahashi-Matsuyama heuristic [5, 6], and able to build a 2-aprroximated Steiner tree. Then, in [7] the Virtual-Source Capacity-Priority algorithm is proposed to reduce the total number of wavelength channels used in the multicast forest, which is viewed as an enhancement of the Member-Only algorithm [1,8]. This enhancement is based on the capabilities of Virtual Source (VS) nodes having both splitting and wavelength conversion capacities. And in [9] the avoidance of *Multicast Incapable Branching* (MIB) nodes multicast routing algorithm is proposed to reduce the average delay and the diameter of the multicast light-trees while obtaining a better cost and link stress than Reroute-to-Any algorithm [1]. However, compared to Member-Only its performance in terms of wavelength channels required and link stress is not favorable.

As time sensitive multicast applications such as video conference, VOIP and On-line games become popular, the average end-to-end delay should be carefully treated, which is an importance parameter for QoS. Meanwhile, the diameter of the multicast tree should also be considered in order to reduce the maximum end-to-end delay as well as the number of costly amplifiers in the tree, which are used to compensate the power loss due to power splitters and distance attenuation. The common problems of the algorithms discussed earlier are the diameter of *light-trees* is too big and average end-to-end delay is also too large, which are not tolerant for QoS required multicast applications. That is why a distance based priority mechanism is proposed for the construction of multicast *light-trees* in this paper. Its significance lays at greatly reducing the diameter of *light-trees* and the average end-to-end delay while keeping the same or even getting a slightly better link stress as well as the number of wavelength channels required than Member-Only.

The rest of the paper is organized as follows. Firstly, the multicast routing problem under optical constraints is discussed in Section II. Then a distance priority based multicast routing algorithm is presented in Section III. In succession, the performance and the simulation of the proposed algorithm are analyzed in Section IV. Finally, a summary is made in section V.

## II. MULTICAST ROUTING IN WDM NETWORKS UNDER SPARSE SPLITTING CONSTRAINT

### A. Multicast Routing Problem

In a WDM optical network, a multicast session $m = \{source = s \mid destinations: d_1, d_2... d_n\}$ is assumed to be required. In order to accommodate the multicast group, a multicast tree or multicast forest should be build to optimize the network resources such as wavelength channels and the number of wavelength used. Furthermore, considering the time sensitive applications and the number of amplifiers required in the multicast tree, the average end-to-end delay and the diameter of the multicast tree should also be taken into account of. As a result, the average end-to-end delay and the diameter of *light-trees* also should be reduced and optimized while achieve the best networks resources utilization.

We assume the wavelength conversion is not available in our WDM optical networks. And, the nodes with splitting capability are also sparse because of their complicate architecture and expensive cost. Hence, there are only two kinds of nodes in optical networks: multicast incapable nodes, multicast capable nodes. Without lack of generality, the splitting capability of multicast capable nodes is assumed to be infinite, which is a very ideal situation. In addition, the hop counts are used as a metric to calculate the cost as well as the delay.

### B. Some Definitions

In order to facilitate later descriptions, some necessary definitions are introduced below.

*Definition 1*: MI and MC nodes

*MI* nodes: Multicast incapable nodes are nodes which can't split, but have Dac capability. That is to say, it can tap a small amount of optical power from the wavelength channel while forward it to only one output link.

*MC* nodes: Multicast capable nodes are nodes which are capable of splitting the incoming message to all of the outgoing ports.

*Definition 2*: Set *MC_SET*, *MI_SET* and ***D***

For a multicast tree,

*MC_SET*: includes the multicast capable nodes (*MC* node) and the leaf multicast incapable nodes (leaf *MI* nodes). They may be used to span the multicast tree.

*MI_SET*: includes only the non-leaf multicast incapable nodes, which are not able to connect a new destination to the multicast tree.

*D*: includes unvisited multicast destinations which are not yet joined to the multicast tree.

*Definition 3*: Constraint Path (*CP*) and Shortest Constraint Path (*SCP*)

A constraint path between a node *u* and a tree *T* is a shortest path from node *u* to a node *v* in the *MC_SET* for *T*, and this shortest path should not traverse any node in *MI_SET* for *T*. That is:

$$CP(u,T) = \{p(u,v) | v \in T \ \& \ \forall x \in p(u,v), x \notin MI\_SET\} \quad (1)$$

where $p(u, v)$ denotes the shortest path from *u* to *v* in the graph. And the constraint path with the minimum length is called the *Shortest Constraint Path* (*SCP*).

$$min\{dist[CP(u,T)]\} = dist\{SCP(u,T)\} \quad (2)$$

Accordingly, node *v* is called the connector for *u* to *T*. There may be several SCPs from *u* to *T* with the same length and so do the connectors.

### C. Member-Only Heuristic

Member-Only algorithm takes account of the splitting constraints of optical nodes when using the Minimum Path Heuristic [8] to build the multicast *light-trees*. The construction of multicast *light-trees* begins with the source: $T = \{s\}$, $MI\_SET = \emptyset$, $MC\_SET = \{s\}$ and $D = \{$all the destinations$\}$. At each iteration, the nearest destination $d_i$ to the nodes in *MC_SET* is added to *T* using the shortest path. But this shortest path should not involve any *MI* nodes in *MI_SET*. It is referred as its Shortest Constraint Path ( $SCP(d_i, T)$ ). That is to say, the length of SCP from $d_i$ to the *T* is the smallest among all the destinations in *D*, which are not yet jointed to the *T*. Then, along the path $SCP(d_i, T)$, update the sets information, which include: (i) add this path to the subtree *T* (ii) add all the *MC* nodes and the leaf *MI* node to *MC_SET*, and delete the non-leaf MI node from *MC_SET*; (iii) add all the non-leaf *MI* nodes to *MI_SET*; (iv) delete the nearest destination $d_i$ from *D*. In succession, in the new set *D*, still find the nearest destination to the new subtree *T* and continue the same procedure as before. If no such destination and *SCP* can be found, then begin with a new tree $T = \{s\}$ until *D* is empty.

## III. DISTANCE PRIORITY BASED MULTICAST ROUTING

### A. Problem Description

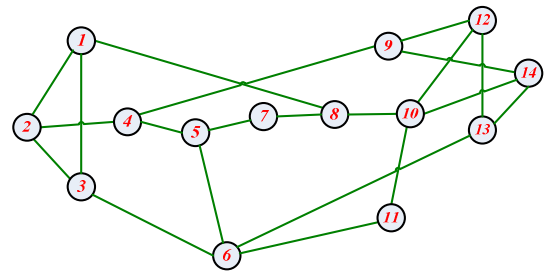

Figure 1. NSF Network

For instance, a multicast session $m_1$ = {source: 2| members: 1~12} is required in the NSF network in Figure 1. We assume, only the source is a *MC* node, and the others are MI nodes. By using Member-Only algorithm, we may get a multicast *light-tree* like in Figure 2(a). As we can see from the multicast *light-tree*, its diameter is 8 and the average delay is 38/11. What is worth noting that its diameter and average delay can be improved by joining node 6 to the tree via node 3 (as shown in Figure 2(a)), while the link stress and the wavelength channels used maintain the same values. From this example, we can see although the best resources utilization can be achieved by using Member-Only algorithm, the diameter of *light-trees* and the average end-to-end delay still could be greatly improved at the same time.

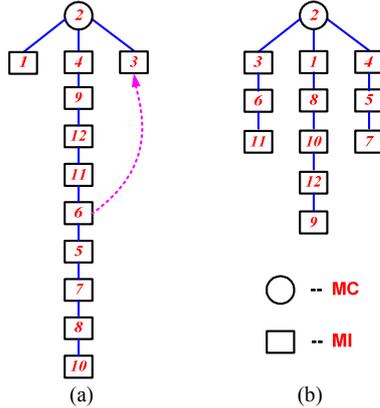

Figure 2. (a) The *light-tree* built using Member-Only. (b) The *light-tree* built by Distance Priority based Algorithm

But, why this situation occurs? In Member-Only algorithm, only one destination is added to the subtree at each step. In fact, it is not difficult to find that there are always several nearest destinations with an equal shortest distance to the subtree at each step. That means not only the lengths of their *SCPs* to the subtree are the same, but also their lengths are the minimum among the destinations in set *D*. Hence, we call them as ***Candidate Destination*** nodes. The problem is, in Member-Only algorithm, any one of them can be selected to join the subtree. However, if the choosing order is different, the result can be totally different. In addition, we can also note that in one step, for one nearest destination *d*, it may have different *SCPs* to the subtree. That is to say, there are several nodes in *MC_SET* for the subtree, say $c_1$ and $c_2$. Both the shortest paths from $c_1$ and $c_2$ to *d* have the shortest length among all the nodes in *MC_SET*, and don't involve any node in *MI_SET* for the subtree. Here, the nodes like $c_1$ and $c_2$ are named as ***Candidate Connector*** nodes in the subtree. In Member-Only, *d* can be joined to the subtree via any one of them. However, if the connector node for *d* is not carefully chosen, then the delay from *d* to the source will not be favorable.

B. *Proposed Algorithm*

In order to reduce the diameter of multicast tree and the average delay, the distance priority heuristic is proposed in the

### Distance Priority Based Multicast Routing Algorithm

***Step 1***: Associate all the destinations with priorities.
  (a) use Dijkstra to compute the *SPT* rooted at source for the multicast session *m*
  (b) assign priorities to destinations in the ascending order of $dist(d_i, s)$ in *SPT* : the nearer, the higher.

***Step 2***: Initialize a tree with the source:
  $T = \{s\}$, $MC\_SET = \{s\}$, $MI\_SET = \emptyset$.

***Step 3***: Find the optimal *Candidate Destination* node *d*.
  (a) compute the $SCP(d_i, T)$ for all $d_i \in D$
  (b) remark destinations whose *SCPs* have equal minimum length as *Candidate Destinations*:
    $dist\{SCP(d, T)\} = \min\{dist[SCP(d_i, T)]|d_i \in D\}$
  (c) select the one with the highest priority as *d*.

***Step 4***: For *d*, find the optimal *Candidate Connector* node *c* in *MC_SET*.
  (a) compute the shortest path $SP(d, connector_i)$ for all $connector_i \in MC\_SET$
  (b) remark the connector nodes whose *SPs* don't traverse any node in *MI_SET* and satisfy (*) as candidate connectors:
    $dist\{SP(d, connector_i)\} = dist\{SCP(d, T)\}$ -- (*)
  (c) among the *Candidate Connector* nodes, choose the one nearest to *s* in *T* as *c* (with the highest priority).

***Step 5***: Connect *d* to *T* using $SP(d, c)$, and update sets.
  *MC_SET*: add *d* and all MC nodes on $SP(d, c)$, remove node *c* if it is a MI node.
  *MI_SET*: add all non-leaf MI nodes on $SP(d, c)$.
  *D*: remove *d*.

***Step 6***: Go to ***step 3*** until no candidate destination node could be found and added to *T*.

***Step 7***: Otherwise, go to ***step2***, until *D* is empty.

TABLE 1.
THE PROCEDURE OF DISTANCE PRIORITY BASED ALGORITHM TO CONSTRUCT THE LIGHT TREE FOR MULTICAST SESSION $m_1$

| Candidate Destinations* | Candidate Connectors[+] | Destinations in D |
|---|---|---|
| ①, 3, 4 | ② | 1, 3~12 |
| ③, 4, 8 | 1, ② | 3~12 |
| ④, 6, 8 | ② | 4~12 |
| ⑤, 6, 8, 9 | ④ | 5~12 |
| ⑥, 7, 8 | ③, 5 | 6~12 |
| 7, ⑧, 11 | ① | 8~12 |
| ⑦, 10, 11 | ⑤, 8 | 7, 9~12 |
| ⑩, 11 | ⑧ | 9~12 |
| ⑪, 12 | ⑥, 10 | 9, 11, 12 |
| ⑫ | ⑩ | 9, 12 |
| ⑨ | ⑫ | 9 |

*① denotes the destination chosen with highest priority
[+]② denotes the connector selected for the destination chosen

construction of multicast tree. This heuristic includes two different standards of priorities. *Candidate Destination* nodes are assigned with different priorities according to their distances to the source in the shortest path tree rooted at the source: the nearer, the higher. While the *Candidate Connector* nodes are offered priorities decided by the distances from them to the source in the subtree being constructed: the nearer, the higher also. The main idea of this heuristic is based on the following two observations (i) if the *Candidate Destination* node with the highest priority is added to the subtree earlier than the others, the diameter of multicast tree could be greatly reduced; (ii) if the nearest destination node is jointed to the subtree via the *Candidate Connector* node with higher priority, then the end-to-end delay from it to the source could also be reduced.

Still see the previous example, through the distance priority heuristic, we can obtain a new multicast tree in Figure 2(b) for the session $m_1$, following steps in Table 1. The diameter of this tree is only 5 and the average delay is 27/11. And its link stress is still 1 and the number of wavelength channels used is still 11, both of which are the same as the multicast tree built by Member-Only algorithm. In fact, if the order for the *Candidate Destination* nodes with the same priority to be added to the subtree is well organized, the result can be even better, which can get a tree with diameter of 4 and average delay of 26/11.

In fact, it is interesting to note that the Minimum Path Heuristic [5] is still respected in our algorithm, which is the key of Member-Only. That is why the link stress and the total number of wavelength channels required could be guaranteed.

## IV. PERFORMANCE EVALUATION AND SIMULATION

In this section we compare the proposed distance priority based algorithm (*DP*) with the famous Member-Only algorithm (*MO*). In[1], Xijun Zhang has showed that Member-Only algorithm provides the best link stress and wavelength usage in the construction of multicast tree or forest. In comparison, four metrics are considered: the diameter of the tree, average delay, link stress as well as the total cost.

The diameter of the tree is defined as the maximum hop counts from each destination to the source in the tree.

$$\textbf{\textit{Diameter}} = max\{dist(d_i, s)| d_i \in m\} \quad (3)$$

**Reduction of Diameter** =
*Diameter*(*MO*) – *Diameter*(*DP*)

The average delay is the average of the end-to-end delays from the source to all the destinations in a multicast session. It is calculated by the sum of the hop counts from each destination to the source divided by the number of destinations.

$$\textbf{\textit{Average Delay}} = \sum_{1 \leq i \leq n} dist(d_i, s)/n \quad (4)$$

**Reduction of Average Delay** =
*Average Delay* (*MO*) – *Average Delay* (*DP*)

, where *n* is the number of destinations in the multicast session.

The link stress is the maximum stress of links in the forest, which equals to the number of wavelengths required.

$$\textbf{\textit{Link Stress}} = \textit{maximum number of wavelengths} \quad (5)$$
*required in one fiber*

The total cost is used to measure the number of wavelength channels used in the multicast tree (forest). It is calculated by the total hop counts in the tree or forest.

$$\textbf{\textit{Total Cost}} = \textit{sum of the hop counts in the} \quad (6)$$
*light tree or forest*

To well evaluate the performance of the proposed algorithm, our simulations are done in both European cost-239 topology (11 nodes) and USA Longhaul topology (28 nodes). The members of a multicast session and the MC nodes in the networks are assumed to be uniformly distributed. Each node in the network has been selected as the source of the multicast session in sequence. For each source, 100 random multicast sessions are generated. Hence, the result of each point in the curves is the average of 100×*number of nodes in the network* computations.

In Figure 3, the results of the simulation in European cost-239 network are presented. The following situations are considered: when the number of members in the multicast session is respectively 4, 6, 8, 10 and 11. As shown in Figure 3(a), (b), we can see the diameter of tree could be reduced up to 2 hop counts (about 45%, (MO-DP)/MO) using the distance based algorithm. And the average delay could also be reduced up to 0.47 hop counts (about 23%). In addition, when the number of members in the session becomes larger, both the reduction of the diameter of the tree and the average delay turn even more significant. What should be noted that, in Figure 3(c)(d), the link stress and the total cost of distance based algorithm are fortunately always the same as those of Member-Only algorithm in all situations. In our simulation, it is also interesting to find that when the number of members in the multicast session is 10 or 11, the average delay can nearly reach the same values as the Reroute-to-Source [1], which is proved to have the optimal delay.

The results of simulation in American Longhaul topology are also presented in Figure 4 in these situations: the number of members in the multicast session equals to 7, 14, 21, 27 and 28. In Figure 4(a)(b), up to 51% (6.75/13.1786) of reduction for the diameter of the multicast tree constructed by our distance based algorithm is found and 50% (3.619/7.2169) for the average delay, while keeping the same links stress and the total number of wavelength channels required. As shown in Figure 4(c)(d), when the number of members is 7 and 14, our proposed algorithm can obtain a slightly smaller link stress than Member-Only algorithm, which is currently thought to have the best link stress. When the number of members is 21, the link stress and the wavelength channels of distance based algorithm are a little bigger than that of Member-Only if the number of MC nodes is smaller than 10. After 10, they become the same. When the number of members is 27 and 28, if the number of MC nodes grows bigger than 7 (That is 25% of nodes are MC nodes), the distance based algorithm gets a slightly better total cost than Member-Only. In this topology, we can also find that the reduction of the diameter of tree and the average delay becomes larger while the number of members in the multicast session grows.

When the number of members in a multicast session is small, neither are there so many *Candidate Destination* nodes for selection at each step, nor are there so many *Candidate Connector* nodes to be found for the *Candidate Destination* with the highest priority. In other word, there are not enough candidates to implement our priorities mechanism. Hence, the advantage of distance priority based algorithm is not so significant. However, when the number of members grows bigger, Member-Only algorithm still choose the candidate nodes arbitrary. In contrary, the proposed algorithm is able to well organize the order of joining to the multicast tree for the *Candidate Destinations* and carefully select the connector nodes. So, in this situation, it is more adorable.

## V. CONCLUSION

Due to physical constraints, supporting multicast routing in optical networks with spare light splitting and without wavelength conversion is not easy. The well-known Member-Only algorithm is currently thought to have the best link stress and the lowest wavelength channels usage. However, its performance in terms of the diameter of the multicast *light-trees* and the average end-to-end delay is not favorable. From this point of view, the distance priority heuristic combined with the minimum cost heuristic is proposed to reduce these two parameters while maintaining the same link stress and the wavelength usage as Member-Only algorithm. Furthermore, it will not produce extra time complexity for the computation of *light-trees* than Member-Only. Simulations have verified that our proposed algorithm is good at reducing the diameter of tree and the average delay in both European Cost-239 network and American Longhaul topology. In fact, similar results could also be verified in the famous NSF network in Figure 2. So, the proposed distance priority based algorithm could be considered as a good candidate algorithm for multicast routing in all optical WDM networks.

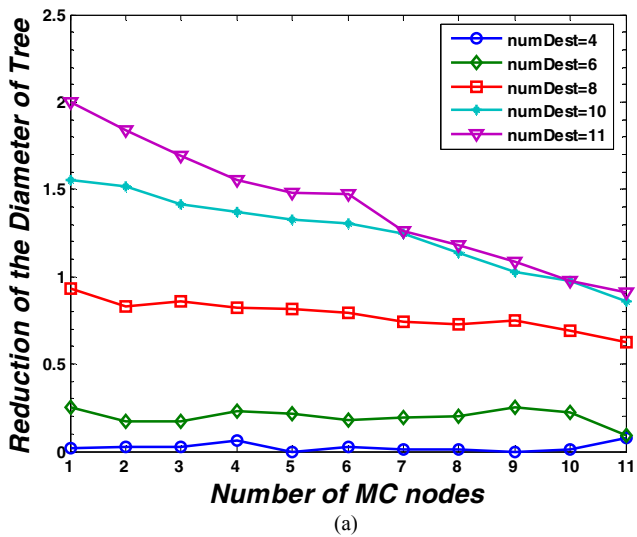
(a)

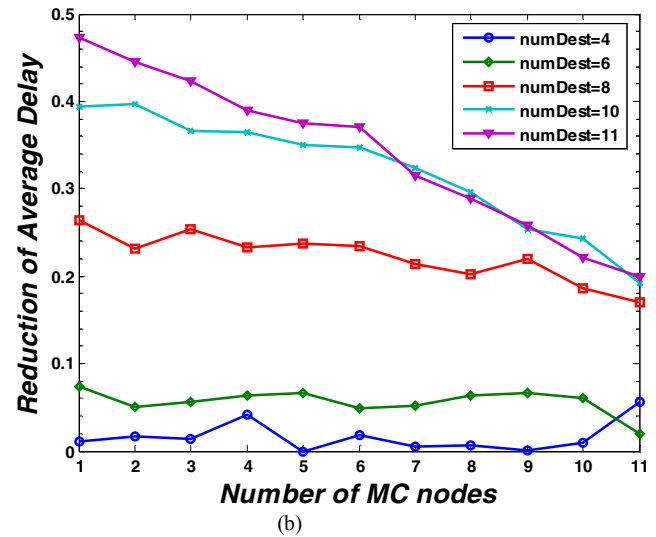
(b)

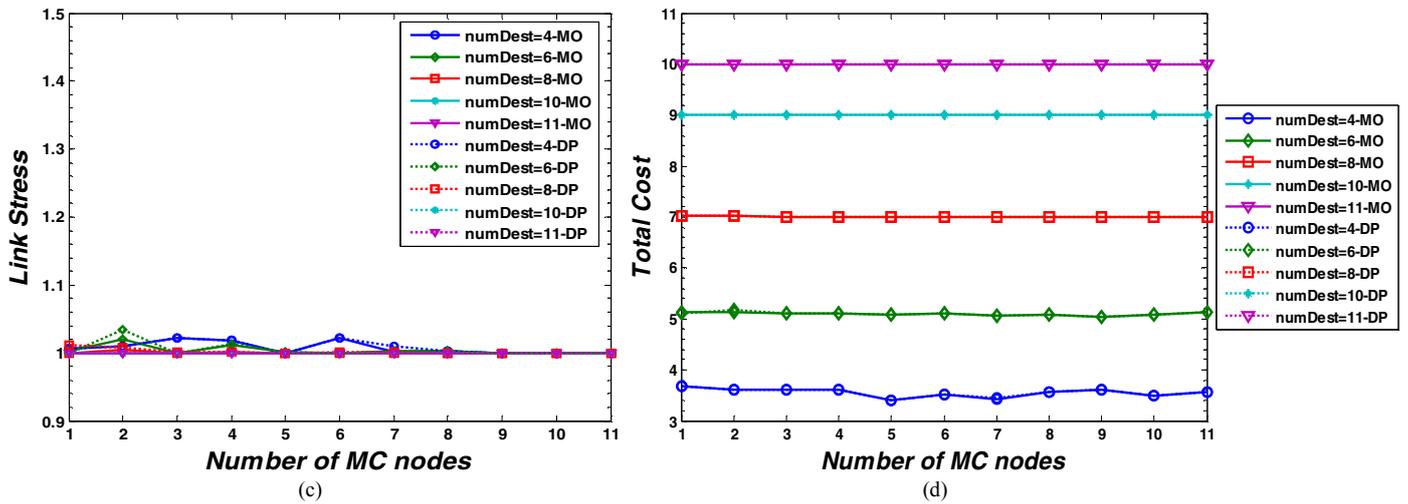

Figure 3. In Europe Cost-239 Topology: (a) Reduction of the Diameter of *light-tree*. (b) Reduction of the Average Delay. (c) Comparison of Link Stress. (d) Comparison of Total Cost.

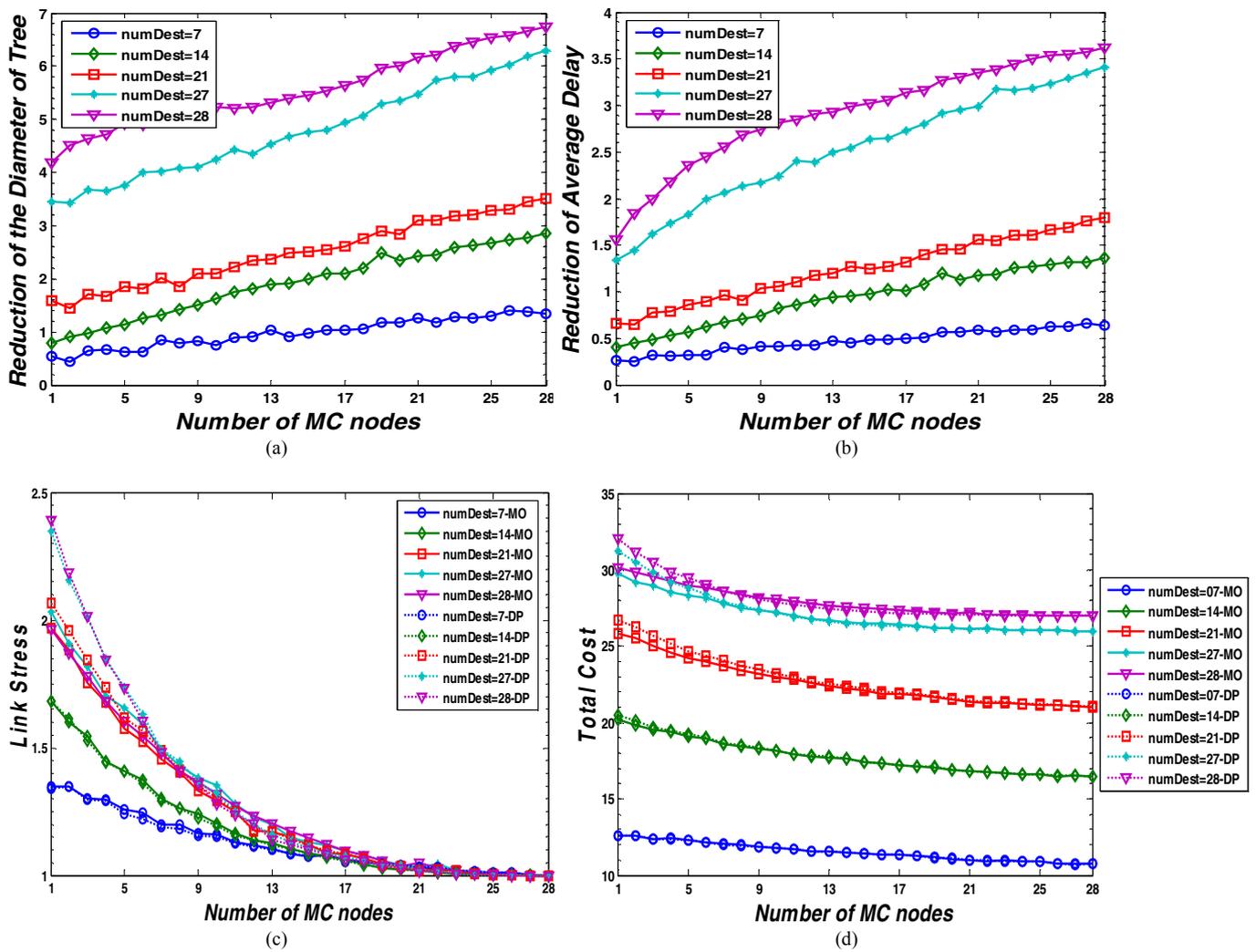

Figure 4. In American Longhaul Topology: (a) Reduction of the Diameter of *light-tree*. (b) Reduction of the Average Delay. (c) Comparison of Link Stress. (d) Comparison of Total Cost.